\documentclass[11pt]{article}
\usepackage{footnote}
\usepackage[dvips]{graphicx}
\usepackage{epsfig}
\usepackage{amssymb}
\usepackage{color}
\usepackage[left]{lineno}
\usepackage{physics}
\usepackage{float}
\usepackage{enumitem}
\usepackage[colorinlistoftodos,prependcaption]{todonotes}
\usepackage{multirow}
\usepackage[alsoload=hep]{siunitx}
\usepackage{pifont}
\usepackage{subfig}
\usepackage{hyperref}
\usepackage[numbers,sort&compress,square]{natbib}

\usepackage{textgreek}

\usepackage[a4paper]{geometry}

\newcommand{\refsec}[1]{Section~\ref{sec:#1}}
\newcommand{\reffig}[1]{Fig.~\ref{fig:#1}}
\newcommand{\reftab}[1]{Table~\ref{tab:#1}}

\newcommand{\labelsec}[1]{\label{sec:#1}}
\newcommand{\labelfig}[1]{\label{fig:#1}}

\newcommand{\gevperc}{\giga\eV\per\clight}

\newcommand{\kpid}{$K_{2\pi D}$ }

\newcommand{\pizero}{\pi^{0}}
\newcommand{\pizerod}{\pi^{0}_{D}}

\DeclareSIUnit\eVperc{\eV\per\clight}
\DeclareSIUnit[number-unit-product = {}]
\percent{\%}
\DeclareSIUnit{\nothing}{\relax}
\DeclareSIUnit{\XO}{\textit{X_0}}

\graphicspath{{./figures/}}

\begin{document}
\centerline{\LARGE EUROPEAN ORGANIZATION FOR NUCLEAR RESEARCH}

\vspace{15mm}
{\flushright{
CERN-EP-2016-323 \\
December 28, 2016\\
Revised version: \\
\today \\
}}
\vspace{15mm}

\begin{center}
\boldmath
{\bf {\Large \boldmath{Measurement of the \(\pi^0\) Electromagnetic \\
Transition Form Factor Slope}}}
\unboldmath
\end{center}

\begin{center}
{\Large The NA62 collaboration$\,$\renewcommand{\thefootnote}{\fnsymbol{footnote}}%
\footnotemark[1]\renewcommand{\thefootnote}{\arabic{footnote}}}\\
\end{center}

\begin{abstract}
The NA62 experiment collected a large sample of charged kaon decays in 2007 with a highly efficient trigger for decays into electrons.
A measurement of the \(\pizero\) electromagnetic transition form factor slope parameter from $1.11\times10^{6}$ fully reconstructed \(K^\pm \to \pi^\pm \pi^0_D, \; \pi^0_D \to e^+ e^- \, \gamma \) events is reported.
The measured value \(a = \qty(3.68 \pm 0.57)\times 10^{-2}\) is in good agreement with theoretical expectations and previous measurements, and represents the most precise experimental determination of the slope in the time-like momentum transfer region.
\end{abstract}

\noindent
\begin{center}
\it{Accepted for publication in Physics Letters B}
\end{center}

\setcounter{footnote}{0}
\renewcommand{\thefootnote}{\fnsymbol{footnote}}
\footnotetext[1]{
\noindent contact: michal.koval@cern.ch, nicolas.lurkin@cern.ch}
\renewcommand{\thefootnote}{\arabic{footnote}}

\newpage

%% restore when needed
\begin{center}
{\Large The NA62 collaboration}\\
\vspace{2mm}
 C.~Lazzeroni$\,$\footnotemark[1],
 N.~Lurkin$\renewcommand{\thefootnote}{\fnsymbol{footnote}}%
\footnotemark[1]^,$\footnotemark[2],
 A.~Romano\\
{\em \small University of Birmingham, Edgbaston, Birmingham, B15 2TT, United Kingdom} \\[0.2cm]
 T.~Blazek, 
 M.~Koval$\renewcommand{\thefootnote}{\fnsymbol{footnote}}%
\footnotemark[1]^,$\footnotemark[3]\\
{\em \small Faculty of Mathematics, Physics and Informatics, Comenius University
in Bratislava, \\ %Mlynska dolina, 
842 48 Bratislava, Slovakia} \\[0.2cm]
 A.~Ceccucci,
 H.~Danielsson,
 V.~Falaleev,
 L.~Gatignon,
 S.~Goy Lopez$\,$\footnotemark[4], \\
 B.~Hallgren$\,$\footnotemark[5],
 A.~Maier,
 A.~Peters,
 M.~Piccini$\,$\footnotemark[6],
 P.~Riedler\\
{\em \small CERN, CH-1211 Gen\`eve 23, Switzerland} \\[0.2cm]
 P.L.~Frabetti,
 E.~Gersabeck$\,$\footnotemark[7],
 V.~Kekelidze,
 D.~Madigozhin,
 M.~Misheva$\,$\footnotemark[8],\\
 N.~Molokanova,
 S.~Movchan,
 Yu.~Potrebenikov,
 S.~Shkarovskiy,
 A.~Zinchenko$\,$\renewcommand{\thefootnote}{\fnsymbol{footnote}}%
\footnotemark[2]\renewcommand{\thefootnote}{\arabic{footnote}},\\
{\em \small Joint Institute for Nuclear Research, 141980 Dubna (MO), Russia} \\[0.2cm]
 P.~Rubin$\,$\footnotemark[9]\\
{\em \small George Mason University, Fairfax, VA 22030, USA} \\[0.2cm]
 W.~Baldini,
 A.~Cotta Ramusino,
 P.~Dalpiaz,
 M.~Fiorini,
 A.~Gianoli, \\
 A.~Norton,
 F.~Petrucci,
 M.~Savri\'e,
 H.~Wahl\\
{\em \small Dipartimento di Fisica e Scienze della Terra dell'Universit\`a e Sezione
dell'INFN di Ferrara, \\ I-44122 Ferrara, Italy} \\[0.2cm]
 A.~Bizzeti$\,$\footnotemark[10],
 F.~Bucci$\,$\footnotemark[11],
 E.~Iacopini$\,$\footnotemark[11],
 M.~Lenti,
 M.~Veltri$\,$\footnotemark[12]\\
{\em \small Sezione dell'INFN di Firenze, I-50019 Sesto Fiorentino, Italy} \\[0.2cm]
 A.~Antonelli,
 M.~Moulson,
 M.~Raggi$\,$\footnotemark[13],
 T.~Spadaro \\
{\em \small Laboratori Nazionali di Frascati, I-00044 Frascati, Italy}\\[0.2cm]
 K.~Eppard,
 M.~Hita-Hochgesand,
 K.~Kleinknecht,
 B.~Renk,
 R.~Wanke,
 A.~Winhart$\,$\footnotemark[5]\\
{\em \small Institut f\"ur Physik, Universit\"at Mainz, D-55099 Mainz, Germany$\,$\footnotemark[14]} \\[0.2cm]
 R.~Winston\\
{\em \small University of California, Merced, CA 95344, USA} \\[0.2cm]
 V.~Bolotov$\,$\renewcommand{\thefootnote}{\fnsymbol{footnote}}%
\footnotemark[2]\renewcommand{\thefootnote}{\arabic{footnote}},
 V.~Duk$\,$\footnotemark[6],
 E.~Gushchin\\
{\em \small Institute for Nuclear Research, 117312 Moscow, Russia} \\[0.2cm]
 F.~Ambrosino,
 D.~Di Filippo,
 P.~Massarotti,
 M.~Napolitano,
 V.~Palladino$\,$\footnotemark[15],
 G.~Saracino \\
{\em \small Dipartimento di Fisica dell'Universit\`a e Sezione dell'INFN di Napoli, I-80126 Napoli, Italy}\\[0.2cm]
 G.~Anzivino,
 E.~Imbergamo,
 R.~Piandani$\,$\footnotemark[16],
 A.~Sergi$\,$\footnotemark[5]\\
{\em \small Dipartimento di Fisica dell'Universit\`a e Sezione dell'INFN di Perugia, I-06100 Perugia, Italy} \\[0.2cm]
 P.~Cenci,
 M.~Pepe\\
{\em \small Sezione dell'INFN di Perugia, I-06100 Perugia, Italy} \\[0.2cm]
 F.~Costantini,
 N.~Doble,
 S.~Giudici,
 G.~Pierazzini$\,$\renewcommand{\thefootnote}{\fnsymbol{footnote}}%
\footnotemark[2]\renewcommand{\thefootnote}{\arabic{footnote}},
 M.~Sozzi,
 S.~Venditti\\%$\,$\footnotemark[2]\\
{\em Dipartimento di Fisica dell'Universit\`a e Sezione dell'INFN di Pisa, I-56100 Pisa, Italy} \\[0.2cm]
 S.~Balev$\,$\renewcommand{\thefootnote}{\fnsymbol{footnote}}%
\footnotemark[2]\renewcommand{\thefootnote}{\arabic{footnote}},
 G.~Collazuol$\,$\footnotemark[17],
 L.~DiLella,
 S.~Gallorini$\,$\footnotemark[17],
 E.~Goudzovski$\,$\renewcommand{\thefootnote}{\arabic{footnote}}\footnotemark[1]$^,$\footnotemark[2]$^,$\footnotemark[5],
 \\
 G.~Lamanna$\,$\footnotemark[18],
 I.~Mannelli,
 G.~Ruggiero$\,$\footnotemark[19]\\
{\em Scuola Normale Superiore e Sezione dell'INFN di Pisa, I-56100 Pisa, Italy} \\[0.2cm]
 C.~Cerri,
 R.~Fantechi \\
{\em Sezione dell'INFN di Pisa, I-56100 Pisa, Italy} \\[0.2cm]
 S.~Kholodenko,
 V.~Kurshetsov,
 V.~Obraztsov,
 V.~Semenov,
 O.~Yushchenko\\
{\em \small Institute for High Energy Physics, 142281 Protvino (MO), Russia}$\,$\footnotemark[20] \\[0.2cm]
\newpage
 G.~D'Agostini\\
{\em \small Dipartimento di Fisica, Sapienza Universit\`a di Roma and\\Sezione dell'INFN di Roma I, I-00185 Roma, Italy} \\[0.2cm]
 E.~Leonardi,
 M.~Serra,
 P.~Valente\\
{\em \small Sezione dell'INFN di Roma I, I-00185 Roma, Italy} \\[0.2cm]
 A.~Fucci,
 A.~Salamon\\
{\em \small Sezione dell'INFN di Roma Tor Vergata, I-00133 Roma, Italy} \\[0.2cm]
 B.~Bloch-Devaux$\,$\footnotemark[21],
 B.~Peyaud\\
{\em \small DSM/IRFU -- CEA Saclay, F-91191 Gif-sur-Yvette, France} \\[0.2cm]
 J.~Engelfried\\
{\em \small Instituto de F\'isica, Universidad Aut\'onoma de San
Luis Potos\'i, 78240 San Luis Potos\'i, Mexico}$\,$\footnotemark[22] \\[0.2cm]
 D.~Coward\\
{\em \small SLAC National Accelerator Laboratory, Stanford University, Menlo Park, CA 94025, USA} \\[0.2cm]
 V.~Kozhuharov$\,$\footnotemark[23],
 L.~Litov \\
{\em \small Faculty of Physics, University of Sofia, 1164 Sofia, Bulgaria}$\,$\footnotemark[24] \\[0.2cm]
 R.~Arcidiacono$\,$\footnotemark[25],
 S.~Bifani$\,$\footnotemark[5] \\
{\em \small Dipartimento di Fisica dell'Universit\`a e
Sezione dell'INFN di Torino, I-10125 Torino, Italy} \\[0.2cm]
 C.~Biino,
 G.~Dellacasa,
 F.~Marchetto \\
{\em \small Sezione dell'INFN di Torino, I-10125 Torino, Italy} \\[0.2cm]
 T.~Numao,
 F.~Reti\`{e}re \\
{\em \small TRIUMF, %4004 Wesbrook Mall, 
Vancouver, British Columbia, V6T 2A3, Canada} \\[0.2cm]
\end{center}
%
%%%%%%%%%%%%%%%%%%%%%%%%%%%%%%%%%
%
\renewcommand{\thefootnote}{\fnsymbol{footnote}}
\footnotetext[1]{Corresponding author, email: nicolas.lurkin@cern.ch, michal.koval@cern.ch}
\footnotetext[2]{Deceased}
\renewcommand{\thefootnote}{\arabic{footnote}}
\footnotetext[1]{Supported by a Royal Society University Research Fellowship}
\footnotetext[2]{Supported by ERC Starting Grant 336581}
\footnotetext[3]{Present address: CERN, CH-1211 Gen\`eve 23, Switzerland}
\footnotetext[4]{Present address: CIEMAT, E-28040 Madrid, Spain}
\footnotetext[5]{Present address: School of Physics and Astronomy, University of Birmingham, Birmingham, B15 2TT, UK}
\footnotetext[6]{Present address: Sezione dell'INFN di Perugia, I-06100 Perugia, Italy}
\footnotetext[7]{Present address: Ruprecht-Karls-Universit\"{a}t Heidelberg, D-69120 Heidelberg, Germany}
\footnotetext[8]{Present address: Institute of Nuclear Research and Nuclear Energy of Bulgarian Academy of Science (INRNE--BAS), Sofia, Bulgaria}
\footnotetext[9]{Funded by the National Science Foundation under award No. 0338597}
\footnotetext[10]{Also at Dipartimento di Fisica, Universit\`a di Modena e Reggio Emilia, I-41125 Modena, Italy}
\footnotetext[11]{Also at Dipartimento di Fisica, Universit\`a di Firenze, I-50019 Sesto Fiorentino, Italy}
\footnotetext[12]{Also at Istituto di Fisica, Universit\`a di Urbino, I-61029 Urbino, Italy}
\footnotetext[13]{Present address: Universit\`a di Roma ``La Sapienza’’, Roma, Italy}
\footnotetext[14]{Funded by the German Federal Minister for Education and Research (BMBF) under contract 05HA6UMA}
\footnotetext[15]{Present address: Physics Department, Imperial College London, London, SW7 2BW, UK}
\footnotetext[16]{Present address: Sezione dell'INFN di Pisa, I-56100 Pisa, Italy}
\footnotetext[17]{Present address: Dipartimento di Fisica dell'Universit\`a e Sezione dell'INFN di Padova, I-35131 Padova, Italy}
\footnotetext[18]{Present address: Dipartimento di Fisica dell'Universit\`a e Sezione dell'INFN di Pisa, I-56100 Pisa, Italy}
\footnotetext[19]{Present address: Department of Physics, University of Liverpool, Liverpool, L69 7ZE, UK}
\footnotetext[20]{Partly funded by the Russian Foundation for Basic Research grant 12-02-91513}
\footnotetext[21]{Present address: Dipartimento di Fisica dell'Universit\`a di Torino, I-10125 Torino, Italy}
\footnotetext[22]{Funded by Consejo Nacional de Ciencia y Tecnolog\'{\i}a {\nobreak (CONACyT)} and Fondo de Apoyo a la Investigaci\'on (UASLP)}
\footnotetext[23]{Also at Laboratori Nazionali di Frascati dell'INFN, Italy}
\footnotetext[24]{Funded by the Bulgarian National Science Fund under contract DID02-22}
\footnotetext[25]{Also at Universit\`a degli Studi del Piemonte Orientale, I-13100 Vercelli, Italy}

\clearpage

%\begin{linenumbers}

%%%%%%%%%%%%%%%%%%%%%%%%%%%%%%%%%%%%%%%
\section*{Introduction}

The Dalitz decay \(\pizerod\to e^{+}e^{-}\gamma\) with a branching fraction of \(\mathcal{B}=\SI{1.174(35)}{\percent}\) \cite{PDG_2016} proceeds through a \(\pizero\to\gamma\gamma^*\) process with an off-shell photon converting into an \(e^+e^-\) pair.
The \(\pi^0\) electromagnetic transition form factor (TFF) describes the deviation of this transition from a point-like interaction. 
It is an input to the computation of the \(\pi^0 \to e^{+} e^{-}\) decay rate \cite{Pi0-ee-Husek_2014}, as well as the hadronic light-by-light scattering contribution to the muon anomalous magnetic moment \((g-2)_\mu\) which at present contributes the second largest uncertainty on its Standard Model value~\cite{nyffeler_2016}.
The commonly used kinematic variables are defined in terms of the \(e^\pm\) and \(\pi^0\) four-momenta (\(p_{e^\pm}\), \(p_{\pi^0}\)) as

\begin{equation*}
    \label{eq:pi0-xydef}
    x = \left( \frac{M_{e e}}{m_{\pi^0}} \right)^2
    = \frac{(p_{e^+} + p_{e^-})^2}{ m_{\pi^0}^2}, \qquad
    y = \frac{2 \, p_{\pi^0} \cdot \left( p_{e^+} - p_{e^-} \right)}{m_{\pi^0}^2
    (1-x)} \;,
\end{equation*}
with the allowed kinematic region defined as

\begin{equation*}
    \label{eq:pi0-xylimits}
    r^2 = \left(\frac{2 m_e}{m_{\pi^0}}\right)^2 \leq x \leq 1, \quad |y| \leq
    \sqrt{1 - \frac{r^2}{x}} \;,
\end{equation*}
where \(m_e\) and \(m_{\pi^0}\) are the corresponding PDG \cite{PDG_2016} masses, and \(M_{ee}\) is the invariant mass of the \(e^+e^-\) pair.
The differential decay width reads \cite{Pi0-Joseph_1960}

\begin{equation*}
    \label{eq:dalitz-dgdxy}
    \frac{\text{d}^2 \Gamma(\pi^0_D)}{\text{d}x \text{d}y} =
    \frac{\alpha}{4 \pi}\Gamma(\pi^0_{2\gamma}) \frac{(1-x)^3}{x} \qty(1 + y^2 +
    \frac{r^2}{x}) \; \qty(1+\delta(x,y)) \; \abs{\mathcal{F}(x)}^2 \;,
\end{equation*}
where \(\Gamma(\pi^0_{2\gamma})\) is the \(\pi^0 \to \gamma\gamma\) decay width, the function \(\delta(x,y)\) describes the radiative corrections and \(\mathcal{F}(x)\) is the electromagnetic transition form factor of the \(\pizero\) to a real and virtual photon.
The function \(\mathcal{F}(x)\) is expected to vary slowly in the kinematic region of the \(\pizero_D\) decay and is usually approximated by a linear expansion \(\mathcal{F}(x) = 1 + a x\), where \(a\) is the \emph{slope} parameter.
%The \(\pi^0\) TFF is an input to the computation of quantities such as the \(\pi^0 \to e^{+} e^{-}\) decay rate \cite{Pi0-ee-Husek_2014} and the anomalous magnetic moment of the muon \cite{mu_g-2-Jegerlehner_2009}.
%The \(\pi^0\) TFF is an input to the computation of quantities such as the \(\pi^0 \to e^{+} e^{-}\) decay rate \cite{Pi0-ee-Husek_2014} and the hadronic light-by-light scattering which is important for the anomalous magnetic moment of the muon \((g-2)_\mu\) \cite{mu_g-2-Jegerlehner_2009}.
The vector meson dominance (VMD) model \cite{Pi0-TFF-Gell-Mann_1961,Pi0-TFF-Lichard_2011} predicts a \(\pizero\) TFF slope value of \(a \approx 0.03\), in agreement with further theoretical estimates \cite{Pi0-radcorr-Kampf_2006, Pi0-TFF-Masjuan_2012, Pi0-TFF-Hoferichter_2014, Pi0-TFF-Husek_2015}.

The TFF slope has been determined in the time-like momentum transfer region by measuring the \(\pizerod\) decay rate \cite{Pi0-TFF-Fischer_1977, Pi0-TFF-Fonvieille_1989, Pi0-TFF-Farzanpay_1992, Pi0-TFF-MeijerDrees_1992, Pi0-TFF-Adlarson_2016}, all including radiative corrections.
The TFF has been measured in the space-like momentum transfer region in the reaction \(e^+e^- \to e^+ e^- \pizero\), where the \(\pizero\) is produced by the fusion of two photons radiated by the incoming beams and decays to two detected photons~\cite{Pi0-TFF-CELLO_1990}. 
The current world average \(a = \num{0.032(4)}\) \cite{PDG_2016} is obtained from time-like measurements \cite{Pi0-TFF-Fonvieille_1989, Pi0-TFF-Farzanpay_1992, Pi0-TFF-MeijerDrees_1992} and the extrapolation of space-like data \cite{Pi0-TFF-CELLO_1990} using a VMD model. 

The NA62 experiment at the CERN SPS collected in 2007 a large sample of charged kaons decaying in flight in vacuum with a minimum-bias trigger configuration \cite{RK-NA62_2013}.
The \(K^\pm\) decays represent a source of tagged neutral pions; the \(K^\pm \to \pi^\pm \pi^0\) (\(K_{2\pi}\)) decay channel accounts for \SI{63}{\percent} of \(\pizero\) production.
The mean free path of the neutral pion in the NA62 experimental conditions is negligible (few \SI{}{\micro\meter}).
This letter reports a model-independent measurement of the \(\pizero\) TFF slope parameter from an analysis of \(\num{1.11e6}~K_{2\pi}\) decays followed by the prompt \(\pizerod\) decay (denoted \(K_{2\pi{}D}\)) using the full NA62 2007 data set.
\section{Beam and detector}
\labelsec{exp_setup}
The NA62 experimental setup used in 2007 was composed of the NA48 detector \cite{NA48_Detector} and a modified beam line \cite{Batley2007} of the earlier NA48/2 experiment.

The beam line was designed to provide simultaneously $K^+$ and $K^-$ beams.
The primary \SI{400}{\gevperc} proton beam delivered by the SPS impinged on a beryllium target of \SI{40}{\cm} length and \SI{0.2}{\cm} diameter.
The secondary beam momenta were selected by magnets in a four dipole achromat and a momentum-defining slit incorporated into a beam dump.
This \SI{3.2}{\m} thick copper/iron block provided the possibility to block either of the \(K^+\) or \(K^-\) beams.
The selected particles had a central momentum of \SI{74}{\gevperc} with a spread of \SI{\pm1.4}{\gevperc} (rms).
The beams were focused and collimated before entering a \SI{114}{\meter} long cylindrical vacuum tank containing the fiducial decay volume.
The beams were mostly composed of \(\pi^\pm\), with a \(K^\pm\) fraction of approximately \SI{6}{\percent}.
Since the muon halo sweeping system was optimised for the positive beam in 2007, most of the data were recorded with the single \(K^+\) beam to reduce the halo background.
The \(K^+\) and \(K^-\) beams were deflected horizontally by a steering magnet at the entrance of the fiducial decay volume at angles of \(\pm\,\SIrange[range-phrase = \text{ to }]{0.23}{0.30}{\milli\rad}\) with respect to the detector axis, to compensate for the opposite \(\mp\, 3.58\,\)mrad deflection by the downstream spectrometer magnet.
The polarities of those magnetic fields were regularly simultaneously reversed to reduce the effects caused by an asymmetry in the detector acceptance.

The momenta of charged particles were measured by a spectrometer composed of four drift chambers (DCH) and a dipole magnet placed between the second and third chamber providing a horizontal transverse momentum kick of \SI{265}{\mega\eVperc} to singly-charged particles.
The measured momentum resolution was \(\sigma_p/p = \SI{0.48}{\percent} \oplus \SI{0.009}{\percent}\cdot p\), where the momentum \(p\) is expressed in \SI{}{\gevperc}.
The spectrometer was housed in a tank filled with helium at nearly atmospheric pressure, separated from the decay volume by a thin (\num{3e-3}\,\(X_0\)) Kevlar\texttrademark{} window.

The photons were detected and measured by a liquid krypton (LKr) electromagnetic calorimeter, which is a quasi-homogeneous ionisation chamber with an active volume of \SI{6.7}{\m\cubed} of octagonal cross-section and a thickness of \SI{127}{\cm}, corresponding to \num{27}\,\(X_0\).
The LKr volume is divided into \SI{13248}{} cells of about \(2\times2\)\SI{}{~\cm\squared} cross section without longitudinal segmentation.
The measured energy resolution was \(\sigma_E/E = \SI{3.2}{\percent}/\sqrt{E} \,\oplus\, \SI{9}{\percent}/E \,\oplus\, \SI{0.42}{\percent}\), and the spatial resolution for the transverse coordinates \(x\) and \(y\) was \(\SI{0.42}{cm}/\sqrt{E} \,\oplus\, \SI{0.06}{\cm}\), where the energy is given in \giga\eV~in both cases.

A scintillator hodoscope (HOD) was located between the spectrometer and the LKr calorimeter.
It consists of a set of scintillators arranged into a plane of 64 vertical counters followed by a plane of 64 horizontal counters. 
Each plane was divided into four quadrants of 16 counters providing a fast trigger signal for charged particles.

%The neutral hodoscope (NHOD) was a vertical plane of scintillating fibers
%installed inside the LKr at a depth of about \SI{9.5}{X_0}. It was used in the trigger to
%provide an independent measurement of the time of the showers and an
%alternative control trigger for neutral particles.

%%%%%%%%%%%%%%%%%%%%%%%%%%%%%%%%
\section{Data sample and trigger logic}
\labelsec{sample_trigger}
% 2x10^10 is taken from the Rk paper
The analysis is based on the full data set collected during 4 months in 2007, corresponding to about \(\num{2e10}\) \(K^\pm\) decays in the vacuum tank.
A total of \SI{65}{\percent} (\SI{8}{\percent}) of the \(K^+\) (\(K^-\)) flux was collected in single-beam mode while the remaining \SI{27}{\percent} were collected with simultaneous \(K^\pm\) beams with a \(K^+/K^-\) flux ratio of \num{2.0}.
During part of the data taking (\SI{55}{\percent} of the \(K^\pm\) flux), a 9.2~\(X_0\) thick transverse horizontal electron absorber lead (Pb) bar was installed between the two HOD planes, approximately 1.2\,m in front of the LKr calorimeter, to study muon-induced electromagnetic showers \cite{RK-NA62_2013}.
A total of 11 rows of LKr calorimeter cells were shadowed by the bar, corresponding to about \SI{10}{\percent} of the total number of cells.

% Rk paper: 10^5 kaon decays in the volume per second
The \SI{100}{\kilo\hertz} kaon decay rate in the vacuum volume during the spill enabled the use of a minimum-bias trigger configuration with a highly efficient trigger chain optimised to select events with at least one electron (\(e^\pm\)) track.

The low level hardware trigger required a coincidence of hits in at least one hodoscope quadrant in both planes (the \(\text{Q}_1\) condition), upper and lower cuts on the hit multiplicity in the drift chambers (the 1-track condition), and a minimum total energy deposit of \SI{10}{\giga\eV} in the LKr calorimeter (the \(E_\text{LKr}\) condition).
The high level software trigger (HLT) condition required at least one track with \(\SI{5}{\gevperc} < p < \SI{90}{\gevperc}\) and \(E/p > 0.6\), where \(E\) is the energy reconstructed in the calorimeter and \(p\) is the momentum reconstructed in the spectrometer.
Downscaled minimum bias trigger streams were collected to evaluate the trigger efficiencies.

%In order to study the trigger efficiencies, the downscaled trigger streams
%\(K_{\mu2}\) (\(\text{Q}_1\) and 1-track only), NHOD (neutral particle trigger using the
%neutral hodoscope), and \(K_{e2}\) autopass (main trigger chain where the
%\(L3(K_{e2})\) condition is only flagged) are also used.

%%%%%%%%%%%%%%%%%%%%%%%%%%%%%%%%
\section{Simulated samples and \texorpdfstring{\mbox{\boldmath$\pizero_D$}}{Pi0D} decay simulation} 
Monte Carlo (MC) simulations of the \(K_{2\pi D}\) decay chain and two other \(K^\pm\) decay chains producing \(\pizero\) Dalitz decays, \(K^{\pm} \to \pi^0_D e^\pm \nu\) and \(K^{\pm} \to \pi^0_D \mu^\pm \nu \) (denoted \(K_{e3D}\) and \(K_{\mu3D}\), respectively), were performed with a \(\pizero\) TFF slope \(a_\text{MC} = \num{3.2e-2}\).
Separate simulated samples, proportionally to the number of kaon decays recorded, were produced for each data taking condition.
The total simulated sample amounts to \SI{386}{M}~\(K_{2\pi D}\), \SI{105}{M} \(K_{\mu3D}\) and \SI{103}{M} \(K_{e3D}\) events within the \SI{97}{\m} long fiducial decay region.
All these modes contribute to the \(\pizerod\) sample, although the selection is optimized for \(K_{2\pi D}\). 

The radiative corrections to the total \cite{Pi0-Joseph_1960} and differential \cite{Pi0-radcorr-Lautrup-Smith_1971, Pi0-radcorr-Mikaelian-Smith_1972, Pi0-radcorr-Husek_2015} \(\pizerod\) decay widths have been studied extensively.
They have to be considered for the TFF measurement since their effect on the \(x\) spectrum is comparable to the effect of the TFF.
The calculation of the radiative corrections \cite{Pi0-radcorr-Husek_2015} implemented in the MC simulation of the \(\pizerod\) decay for the present analysis includes real photon emission from the \(\pizerod\) decay vertex.
It also includes the one-photon irreducible contribution, neglected in earlier studies, which has an effect of \(|\Delta a| \approx \num{0.5e-2}\) on the slope of the \(x\) spectrum.
Higher order correction terms not included in the simulation contribute to the slope by \(|\Delta a| < \num{0.01e-2}\), which is considered as a systematic uncertainty (\reftab{error-budget}).
%

% PHOTOS \cite{Photos-Barberio_1994}

% The previously available radiative correction algorithms in the simulation had
% limitations for this analysis: does not
% reproduce the \(x\) spectrum correctly, and the Mikaelian and Smith
% \cite{Mikaelian_Smith} corrections modifying the LO \(x\) distribution do not
% generate the additional radiative photons. Instead, a new MC \(\pizerod\) event
% generator \cite{Pi0-radcorr-Husek_2015} producing the electron-positron pairs
% according to the correct \(x\) Dalitz shape and extra radiative photon in the
% final state was used. The main idea is to use a cut-off parameter on the
% diphoton invariant mass to switch between the generation of real bremsstrahlung
% photon and the integration of all radiative corrections.

%%%%%%%%%%%%%%%%%%%%%%%%%%%%%%%%%%
\section{Data analysis}

\subsection{Event reconstruction and selection}
\labelsec{selection}

Hits and drift times in the DCH and a detailed map of the magnetic field are used to reconstruct track directions and momenta.
Three-track vertices are reconstructed by a Kalman filter algorithm extrapolating track segments from the upstream part of the spectrometer into the decay volume, taking into account multiple scattering in the helium and the Kevlar window, the Earth's magnetic field and residual vacuum tank magnetization.
The reconstructed \(K^\pm\to\pi^\pm\pi^+\pi^-\) invariant mass and the missing mass in the \(K^\pm\to\mu^\pm \nu\) decay are monitored and used for fine calibration of the spectrometer momentum scale and DCH alignment.
Clusters of energy deposition in the LKr calorimeter are found by locating maxima in space and time in the digitized pulses from individual cells.
Reconstructed energies are corrected for energy outside the cluster boundaries, energy lost in isolated inactive cells (\SI{0.8}{\percent} of the total number), sharing of energy between clusters, and non-linearity for clusters with energy below \SI{11}{\GeV}.
Electrons produced in \(K^\pm\to\pi^0 e^\pm\nu\) decays are used to calibrate the energy response.

The main \(K_{2\pi D}\) selection criteria are the following.
\begin{itemize}
  %############# Vertex
  \item The event should contain exactly one reconstructed 3-track vertex, which should be located within the fiducial decay region and be geometrically compatible with a beam kaon decay.
  The vertex charge \(q_\text{vtx}\), defined as the sum of the track charges, should match the beam charge in the single-beam mode.
  Otherwise it should satisfy a relaxed condition \(|q_\text{vtx}|=1\).
  The track with the charge opposite to \(q_\text{vtx}\) is necessarily an \(e^\pm\) candidate, while the same-sign tracks can be either \(\pi^\pm\) or \(e^\pm\) candidates. 

  %############# Tracks
  \item %The event must not contain any extra track, which could potentially
  %affect the reconstructed \(\pizerod\) events. These are in-time tracks not
  %forming the selected 3-track vertex while being consistent with the
  %hypothesis that they originate from a kaon decay (e.g. an accidental muon
  %from a \(K_{\mu2}\) decay). 
  The tracks are required to be in time (within \SI{25}{\ns} of the trigger time and \SI{15}{\ns} of each other), and within the geometrical acceptance of the drift chambers.
  The allowed track momentum range is \SIrange{2}{74}{\gevperc}, excluding the low momentum range where a \SI{2}{\percent} deficit of data with respect to MC simulation is seen.
  Events with a photon converting into an \(e^+e^-\) pair in the material in or in front of DCH1 (Kevlar window, helium) are suppressed by requiring a minimum distance of \SI{2}{\cm} between the impact points of every track pair in the first drift chamber, as verified by simulation of \(K_{2\pi}\) decays.
  %An inefficiency of the reconstruction for the highly parallel \(e^\pm\)
  %tracks (i.e. with a small opening angle) in the central region of the drift
  %chambers leads to poor data/MC agreement for the electron-positron distance
  %and their impact point radius in DCH1. The agreement is restored by
  %rejecting the event if either of the \(e^\pm\) tracks impact point is
  %located within a \SI{40x40}{\cm} square centered on the beam pipe in DCH1.

  %############ Photon
  \item Reconstructed clusters of energy deposition in the LKr calorimeter are used to identify photon candidates.
  A photon candidate cluster should be geometrically isolated from the track impact points in the LKr calorimeter (\(d_t>\SI{20}{\cm}\) from the same-sign tracks and \(d_t>\SI{10}{\cm}\) from the remaining track), within \SI{10}{\nano\s} of each track and with more than \SI{2}{\giga\eV} of energy.
  The photon 4-momentum is reconstructed assuming that the photon originates from the same vertex as the tracks.
  If more than one photon candidate is found, the event is rejected.
  
  %only one ``good'' cluster per event is permitted. A
  %cluster is considered ``good'' if it is in-time with the vertex and its energy is above
  %\SI{2}{\giga\eV}. The clusters associated with a track in the LKr acceptance,
  %by direct energy deposition or bremsstrahlung are not considered. To avoid
  %any effect related to the Pb wall, clusters located behind it are also
  %ignored when it was present. The photon candidate is reconstructed from the
  %``good'' cluster, assuming that it originates from the same vertex as the
  %tracks. Several geometrical and acceptance cut are applied.

  %############# PID
  \item The total reconstructed momentum should be compatible with the beam momentum, in the range \SIrange{70}{78}{\gevperc}, and there should be no missing transverse momentum with respect to the beam axis within the resolution: \(p_t^2<\SI{e-5}{(\gevperc)\squared}\).
  Particle identification using an \(E/p\) ratio is not required thanks to the low background in the sample, reducing the systematics associated to the misidentification and increasing the \(K_{2\pi D}\) acceptance by more than a factor of two.
  %Because the acceptance of the \(E/p\) based selection peaks at \(x<0.1\) and is already twice smaller at \(x=0.3\), it introduces a bias on the \(x\) distribution shape.
  %A kinematic identification of the event with a more uniform acceptance is used according to the following procedure.
  The \(\pi/e\) ambiguity for the two same-sign tracks is resolved by testing the two possible mass assignments.
  For each hypothesis, the reconstructed kinematic variables should be \(|x|,|y|<1\), and the reconstructed \(e^+e^-\gamma\) and \(\pi^\pm\pi^0\) masses should be close to the nominal ones: \(M_{ee\gamma}\) in the range \SIrange{115}{145}{\mega\eVperc\squared} and \(M_{\pi\pi}\) in the range \SIrange{460}{520}{\mega\eVperc\squared}.
  Only events with a single valid hypothesis are selected.
  The probability of correct (incorrect) mass assignment evaluated with the \(K_{2\pi D}\) MC sample is \SI{99.62}{\percent} (\SI{0.02}{\percent}).
  The remaining \SI{0.36}{\percent} of events have either zero or two valid hypotheses and are rejected.

  %############## Trigger
  \item The trigger conditions described in \refsec{sample_trigger} are reproduced on simulated samples.
  To eliminate edge effects due to different calibration and resolution between the trigger and the offline analysis, tighter variants of the trigger criteria are applied to both data and MC samples.
  The offline \(E_\text{LKr}\) condition requires a minimum of \SI{14}{\giga\eV} of electromagnetic energy in the LKr calorimeter summed over the reconstructed photon and \(e^\pm\) clusters.
  The offline condition corresponding to the HLT requires at least one track whose impact point on the LKr calorimeter front plane is within its acceptance and not behind the Pb bar, \(p>\SI{5.5}{\gevperc}\) and \(E/p > \num{0.8}\), effectively requesting that at least one \(e^\pm\) track is detected in the calorimeter.
  
  %############# Signal region
  \item A \SI{1}{\percent} deficit in the data/MC ratio is seen for events with \(x<0.01\) due to the steeply falling acceptance.
  For this reason the signal region is defined as \(x>0.01\), equivalent to \(M_{ee}> \SI{13.5}{\mega\eV\per\clight\squared}\).
\end{itemize}

The selected \(K_{2\pi D}\) sample amounts to \num{1.11e6} events.
The overall acceptances of the selection evaluated with MC simulations are \SI{1.90}{\percent} for \(K_{2\pi D}\) decays, \SI{0.02}{\percent} for \(K_{\mu3D}\) decays and \SI{0.01}{\percent} for \(K_{e3D}\) decays.
The \(K_{2\pi D}\) acceptances for periods with and without the Pb bar installed are \SI{1.64}{\percent} and \SI{2.23}{\percent}, respectively.

The reconstructed \(\pi^\pm\pi^0\) and \(e^+e^-\gamma\) invariant mass spectra are shown in \reffig{reco-masses}; the mass resolutions obtained from a Gaussian fit are \SI{3.7}{\mega\eVperc\squared} and \SI{1.5}{\mega\eVperc\squared} (rms), respectively.
The reconstructed spectrum of the \(x\) variable and the acceptances for the decay channels considered are shown in \reffig{reco-x-acceptance}.
The \(e^+e^-\) mass resolution determined from the \kpid MC sample can be approximated by \(\sigma_{ee} = \SI{0.9}{\percent}\cdot M_{ee}\), which translates into the resolution on the \(x\) variable as \(\sigma_x = \SI{1.8}{\percent}\cdot x\).

\begin{figure}[ht!]
    \centering
    \subfloat{% \includegraphics[width=0.46\columnwidth]{mk.pdf} }
     \begin{tikzpicture}
 		\draw (0,0) node{\includegraphics[width=0.46\columnwidth]{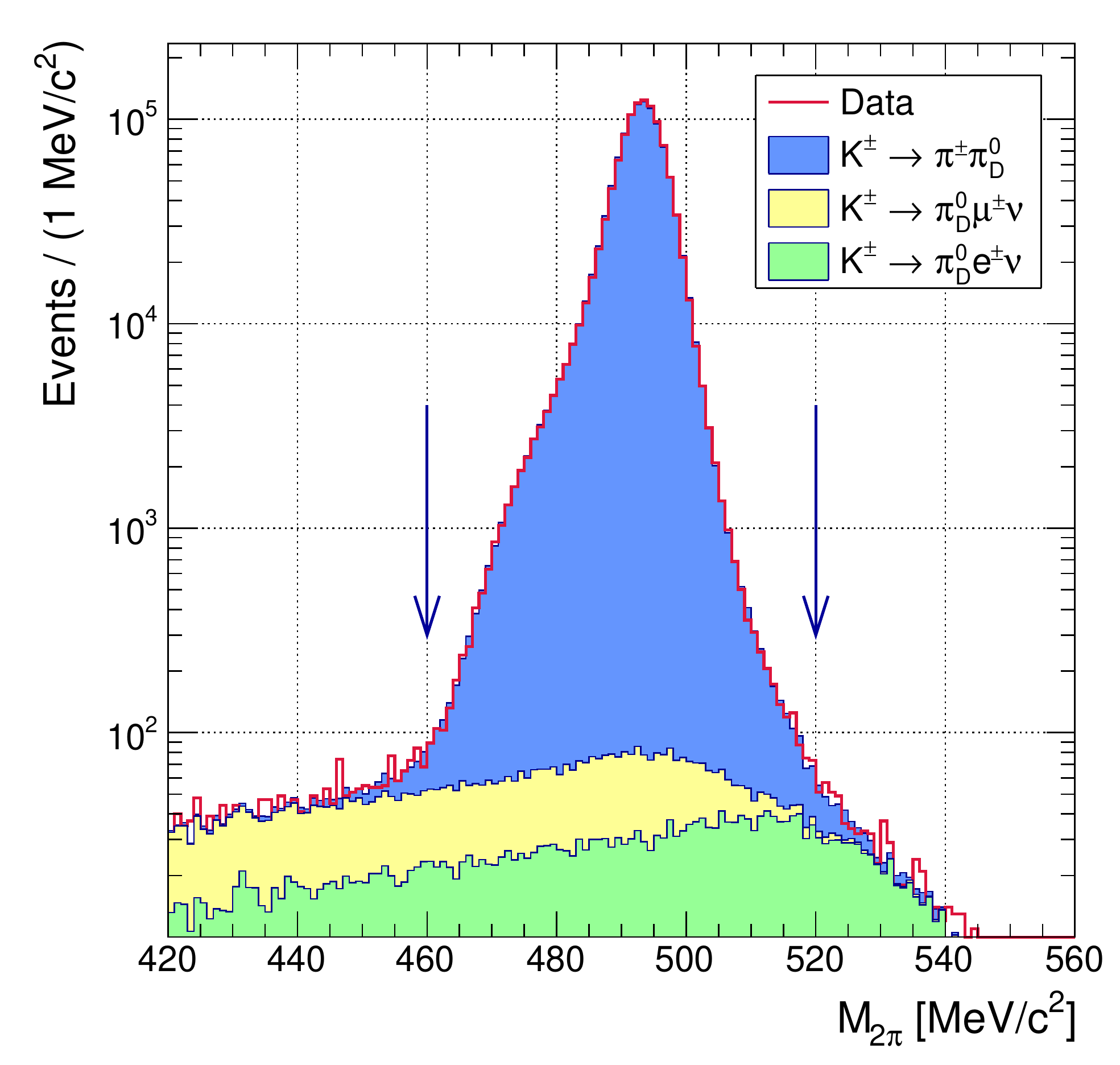}};
 		\draw (-2.1,2.8) node {(a)};
     \end{tikzpicture}
     }
    \subfloat{ %\includegraphics[width=0.46\columnwidth]{mpi0.pdf} }
     \begin{tikzpicture}
 		\draw (0,0) node {\includegraphics[width=0.46\columnwidth]{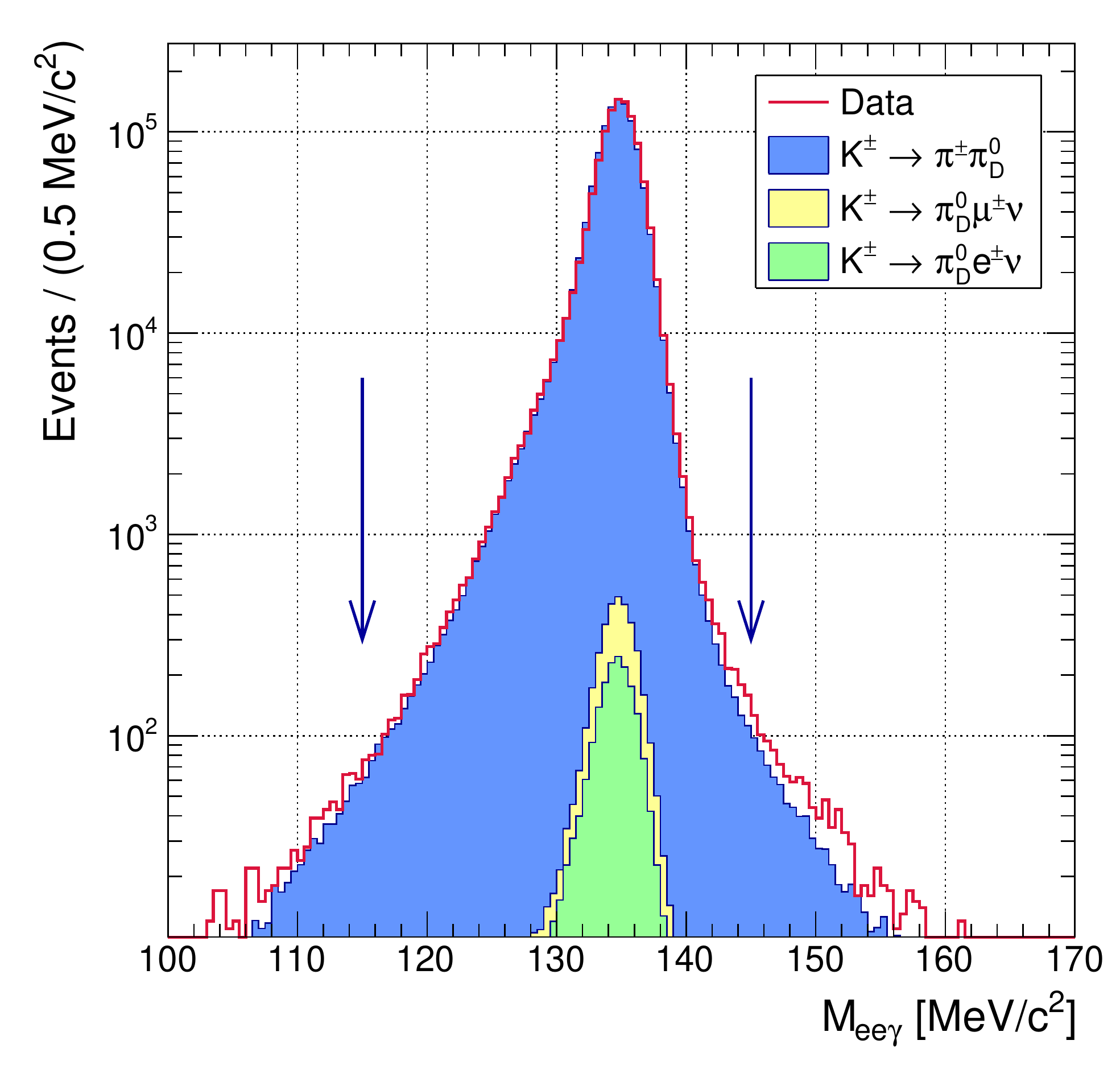}};
     	\draw (-2.1,2.8) node {(b)};
     \end{tikzpicture}
     }
   \caption[Reconstructed invariant mass distributions] {
   Reconstructed (a) \(\pi^\pm\pi^0\) and (b) \(e^+e^-\gamma\) mass distributions for data and simulated components.
   The radiative shoulders in the reconstructed masses are well reproduced in the MC thanks to the simulation of the radiative photon.
   The adopted mass selection criteria represented by the arrows are asymmetric with respect to the nominal \(K^\pm\) and \(\pi^0\) masses.
   }
   \labelfig{reco-masses}
%\end{figure}
%%\vspace{-2em}
%\begin{figure}[ht]
%	\centering
	\subfloat{ %\includegraphics[width=0.48\columnwidth]{xDalitz.pdf}
	\begin{tikzpicture}
		\draw (0,0) node{\includegraphics[width=0.46\columnwidth]{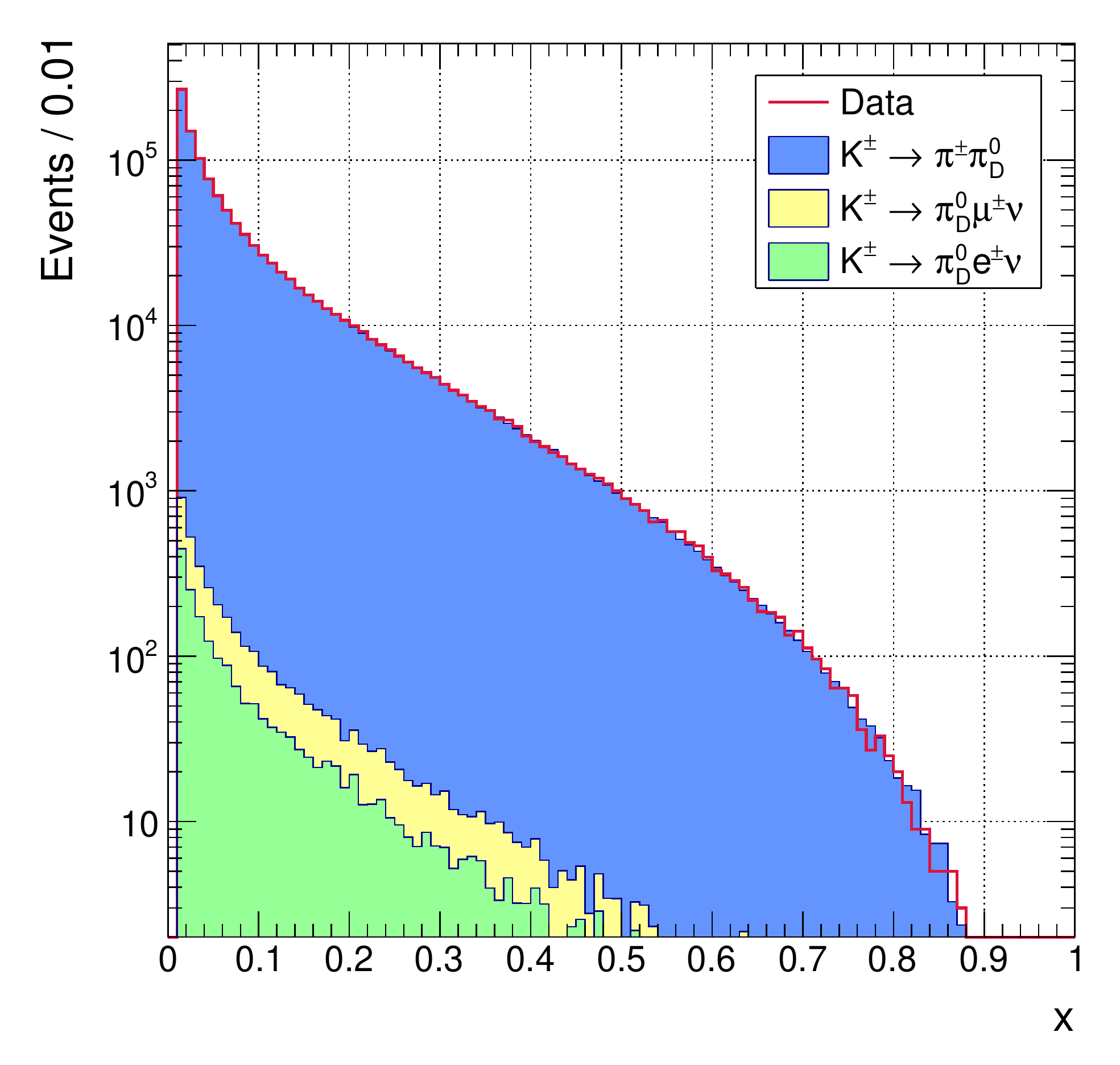}};
		\draw (-2.1,2.8) node {(a)};
    \end{tikzpicture}
	}
	\subfloat{ %\includegraphics[width=0.48\columnwidth]{acceptance_x_mck2pi.pdf}
	\begin{tikzpicture}
		\draw (0,0) node{\includegraphics[width=0.46\columnwidth]{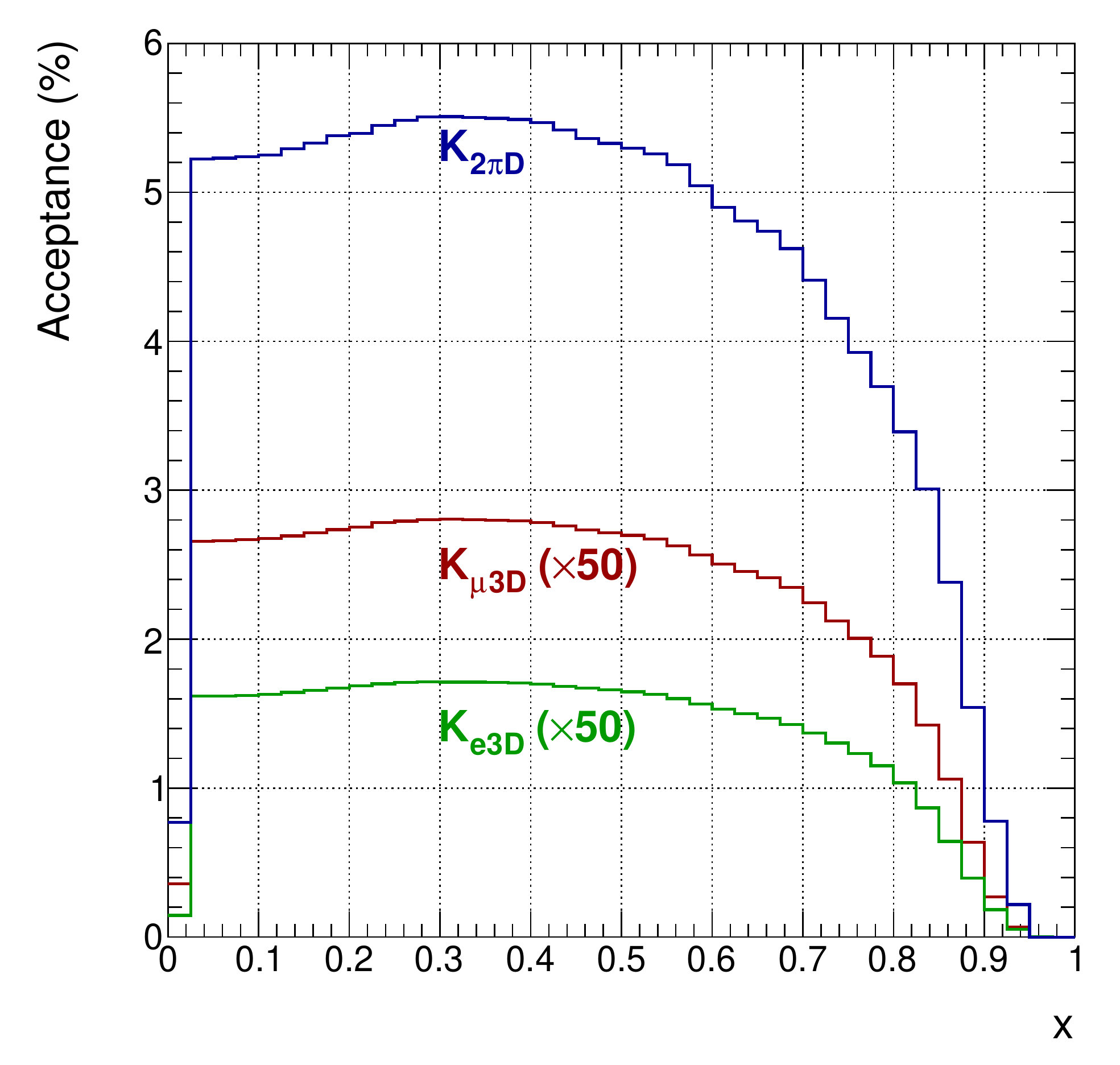}};
		\draw (-2.1,2.8) node {(b)};
    \end{tikzpicture}	
	}
	\caption[Reconstructed $x$ Dalitz variable and acceptance.]
		{(a) Spectra of the reconstructed \(x\) variable for data and MC components. 
		(b) Acceptances of the \(K_{2\pi D}\) selection for each decay as functions of the \(x\) variable.
		The acceptances for \(K_{\ell 3D}\) decays (\(\ell=e;\mu\)) are scaled up by a factor of 50. 
		The drop in the first bin is due to the signal region definition (\(x>0.01\)).}
	\labelfig{reco-x-acceptance}
\end{figure}

%%%%%%%%%%%%%%%%%%%%%%%%%%%%%%%%%%%%%%%%%%%%%%%%%%%

\subsection{Fit procedure}
\labelsec{fit-procedure}
A \(\chi^2\) fit with free MC normalisation in equally populated bins comparing the data and MC reconstructed \(x\) distributions is performed to extract the TFF slope.
A number of slope hypotheses \(a_\text{h}\) are tested by reweighting a single set of MC events simulated with a slope \(a_\text{MC} = \num{3.2e-2}\) using the weights

\begin{equation*}
	w(a_\text{h}) = \frac{(1+a_\text{h}\,x_\text{MC})^2}{(1+a_\text{MC}\,x_\text{MC})^2}\;,
\end{equation*}
where \(x_\text{MC}\) is the true \(x\) value for each event.
The minimization of the \(\chi^2\) test statistics yields the following result:
%A different value of \(a_\text{h}\) is tested at each iteration of the fit until it converges and yields the following result:
 
\begin{equation*}
    a = \left( 3.68 \pm 0.48 \pm 0.18 \right) \times 10^{-2} \;,
\end{equation*}
where the uncertainties are statistical due to the limited data and MC sample sizes.
The fit gives \(\chi^2/\text{ndf} = 54.8/49\), which has a \(p\)-value of \SI{26.4}{\percent}.
The fit result is illustrated in \reffig{fit_illustration}.
Using a quadratic function \(\abs{\mathcal{F}(x)}^2 = 1+2bx+cx^2\), the fit results are \(b = \num{3.71(51)e-2}\) and \(c=\num{0.00(19)}\).

\begin{figure}[ht!]
    \centering
    \includegraphics[width=0.55\columnwidth]{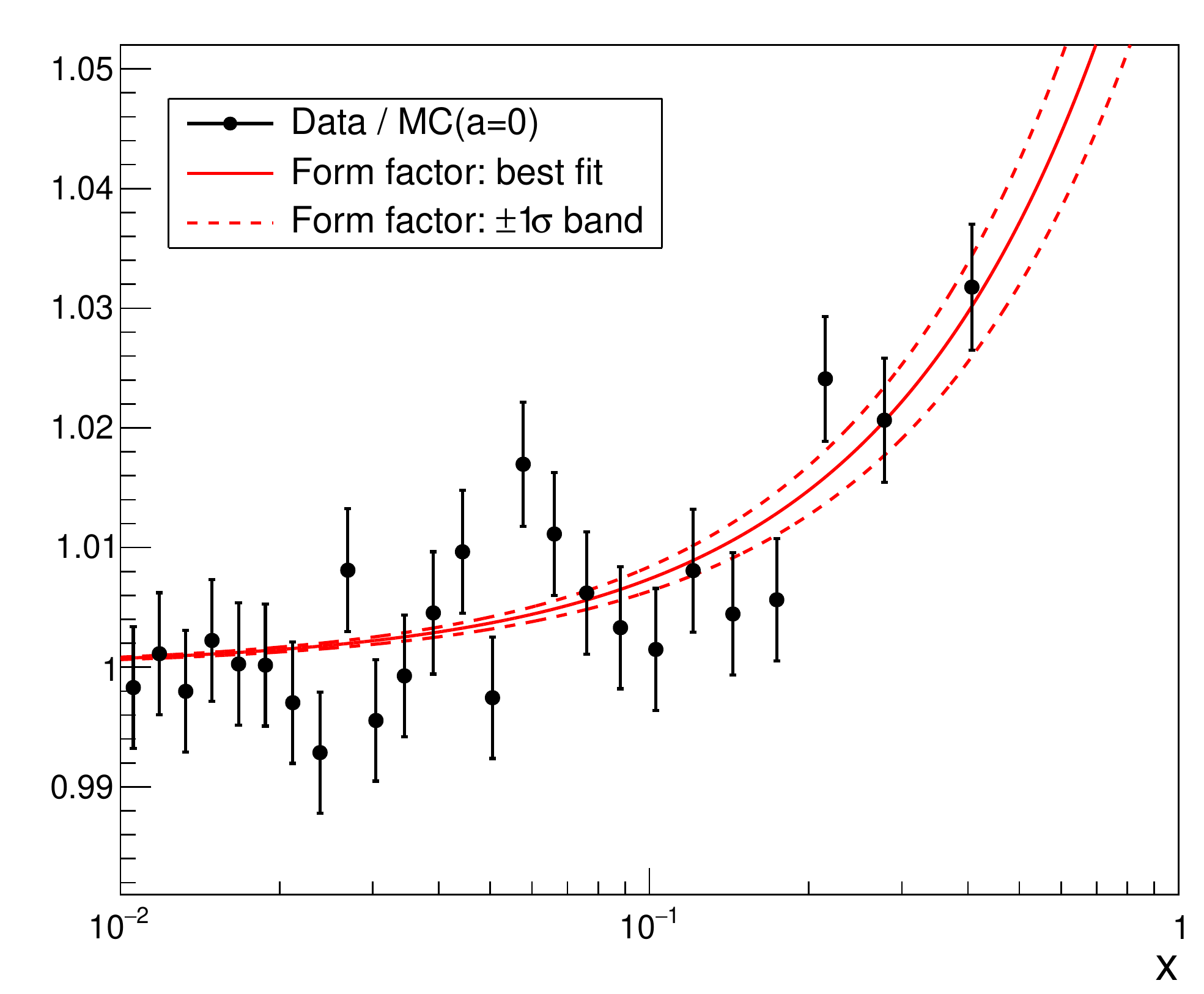}
%    \caption{Illustration of the fit to the TFF. The effect of a
%    positive TFF slope (\(a>0\)) is clearly seen in the ratio of the
%    reconstructed data and MC distributions, with the MC sample weighted to
%    obtain \(a = 0\). Data and MC events are divided into 20 equally populated bins;
%    the horizontal positions of the markers correspond to the bin barycentres.
%    The solid line represents \(|\mathcal{F}(x)|^2\) with the measured slope
%    value. The dashed lines indicate the \(\pm 1 \, \sigma\) band. Only the
%    statistical uncertainties are shown.}
    \caption{Ratio of the reconstructed \(x\) distributions for data and MC, where the MC sample corresponds
    to \(a = 0\). The effect of a positive TFF slope (\(a>0\)) is clearly seen in this illustration.
    Data and MC events are distributed into 25 equally populated bins;
    the horizontal positions of the markers correspond to the bin barycentres.
    The solid line represents \(|\mathcal{F}(x)|^2\) with the measured central slope
    value: \(a = 3.68 \times 10^{-2}\). The dashed lines indicate the \(\pm 1 \, \sigma\) band. Only the
    statistical uncertainties are shown.
    }
    \labelfig{fit_illustration}
\end{figure}

%%%%%%%%%%%%%%%%%%%%%%%%%%%%%%%%%%%%%%%%%%%%%%%%%%%

\subsection{Systematic effects}
\labelsec{syst}

\subsubsection{Calibration, resolution and beam simulation}
\labelsec{calib-resolution}
% The momenta measured in the spectrometer and energy depositions in LKr are
% corrected for several known effects, some of which may affect the measurement:

The spectrometer momentum scale modifies proportionally the \(x\) variable.
%A \SI{1}{\percent} effect on the momentum calibration generates an additional slope of \(|\Delta a| = \num{7.8e-2}\) in the \(x\) distribution.
%Spectrometer momentum scale and chamber alignment corrections modify the reconstructed \(e^\pm\) momenta and thus the \(x\) variable.
%The relative effect of these corrections on the reconstructed momenta is of the order of \num{e-3}, which induces a \SI{4}{\percent} relative change in the slope value.
The corrections applied to the momentum calibration have a typical relative size of the order of \num{e-3}.
The sensitivity of the fit to a residual miscalibration has been assessed conservatively by turning the corrections off, leading to a shift of the fit result of \(\Delta a = \num{-0.16e-2}\) considered as the systematic uncertainty on the spectrometer calibration.
A similar procedure is applied for the chamber misalignment correction with no significant effect on the fit result.

The spectrometer mass resolution has been evaluated separately for individual data-taking periods using samples of \(K^\pm\to\pi^\pm\pi^+\pi^-\) decays.
The maximum relative difference observed on the resolution of the reconstructed squared 3-pions mass between data and MC is \SI{2}{\percent}.
Scaling the MC resolution of the \(x\) variable by \num{0.98} results in a shift of \(\Delta a = \num{0.05e-2}\), which is considered as a systematic uncertainty.

The corrections applied to the energies measured in the LKr calorimeter affect the TFF slope result indirectly through the photon selection acceptance.
A correction for the non-linearity in the energy response in the data sample with an alternative function is used to evaluate the sensitivity to the correction function, resulting in a shift of \(\Delta a = \num{0.03e-2}\).
A global photon energy scaling factor of 1.001, which is the typical size of the energy corrections, applied only in the MC sample causes a shift of \(\Delta a = \num{0.02e-2}\).
The overall systematic uncertainty due to the LKr energy calibration is assigned as the sum of these two effects in quadrature: \(\Delta a = \num{0.04e-2}\).

The beam momentum is simulated according to the central value measured separately for different data taking periods from fully reconstructed \(K^\pm\to\pi^\pm\pi^+\pi^-\) decays.
A remaining discrepancy between data and MC in the tails of the beam momentum spectrum affects the TFF slope measurement through the \(K^\pm\) momentum dependence of the acceptance.
After applying a correction to improve the spectrum data/MC agreement, the measured slope shifts by \(\Delta a = \num{0.03e-2}\), which is considered as a systematic uncertainty.

\subsubsection{Trigger efficiency}
\labelsec{trigger-eff}
%The \(K_{e2}\) trigger chain was used to acquire the present data sample and
%the associated conditions are presented in \refsec{sample}. The stored events
%are tagged with the decision of every trigger line. The efficiency of a criteria
%is tested using events selected by a different trigger streams and not
%containing it.
The efficiencies of individual components of the signal trigger chain have been measured using control data samples collected via alternative trigger chains.
Since no inefficient events have been found, upper limits on the inefficiencies at \SI{90}{\percent} CL have been evaluated for each trigger conditions: \SI{0.06}{\percent} (\(\text{Q}_1\)), \SI{0.10}{\percent} (1-track), \SI{0.03}{\percent} (\(E_\text{LKr}\)) and \SI{0.03}{\percent} (HLT).

Possible systematic effects caused by each trigger condition have been investigated separately by removing potentially inefficient events either from the data or the MC sample.
The \(\text{Q}_1\) efficiency is modeled by introducing fully inefficient gaps between the HOD quadrants with \SI{0.2}{\mm} width tuned using data/MC comparison in other decay channels. 
This leads to a \(\text{Q}_1\) inefficiency of \SI{0.02}{\percent} for \(K_{2\pi D}\) events.
Energetic photons may initiate showers by interacting with the beam pipe material, causing the DCH hit multiplicities to exceed the limits allowed by the 1-track trigger condition.
The sensitivity to this effect is tested by removing from the MC \(K_{2\pi D}\) sample \SI{0.10}{\percent} of events with a radiative photon with an energy above \SI{0.5}{\giga\eV} traversing the beam pipe.
For the \(E_\text{LKr}\) and HLT triggers, events closest to failing a trigger condition are removed from the data sample.
Those are events with the lowest reconstructed energy in the LKr calorimeter for the \(E_\text{LKr}\) condition, and events with the lowest maximum track \(E/p\) ratio for the HLT condition.
In both cases the fraction of removed events is equal to the upper limits on inefficiencies quoted above.
The only sizeable change in the TFF slope result has been observed by testing the \(E_\text{LKr}\) trigger condition, resulting in a systematic uncertainty estimate of \(\Delta a = \num{0.06e-2}\).
 
%\subsubsection{Accidentals and Other
%\texorpdfstring{\mbox{\boldmath$\pizero_D$}}{Pi0D} Sources}
\subsubsection{Backgrounds}

% In the sample of data passing the \(K_{2\pi D}\) selection, there might
% be events which do not come from genuine \(\pizero\) Dalitz decays.

% Those could be \(K^\pm \to \pi^\pm \pi^+ \pi^-\) events with an accidental
% photon in time with the event.

The effect of accidental background is investigated by releasing independently the timing cuts and constraints on the numbers of tracks and vertices in the selection.
The number of additional events included into the data sample for each variation of the selection is less than \num{6e3}.
The total systematic uncertainty due to accidentals is evaluated to be \(\Delta a = \num{0.15e-2}\).

%result is estimated by performing the following modifications to the default
%\(K_{2\pi D}\) selection, described in \refsec{selection}.
%
%A first test is done by releasing the timing cut on the clusters or
%tracks with respect to the trigger time in the data. The additional 0.5\% events
%gained are out of time. The shift \(\Delta a = \num{0.10e-2}\) induced
%on the fit result can be considered an estimate of the effect of accidental events
%possibly present in the default data sample.
%
%The second type of tests examines the influence of accidental or ghost tracks in
%the reconstruction of the event. They can cause the appearance of additional
%vertices, or a change of their total charge. Allowing extra-tracks to be present
%in the event changes the central value of the fit by \(\Delta a =
%\num{-0.06e-2}\). In case multiple vertices are present in the event,
%the whole selection process can be executed on each of them, including in the
%final sample the vertex with best \(\chi^2\) amongst the remaining ones. The
%slope shift obtained when adding these events is \(\Delta a =
%\num{-0.09e-2}\). The number of additional events for each variation
%of the selection is less than 0.5\%.

%\subsubsection{PID, Subsamples}
% Mis-identification of charged tracks can affect the TFF slope fit by introducing
% non \(\pizerod\) background events into the sample or by swapping the
% \(\pi^\pm\) and \(e^\pm\) tracks in the \(K_{2\pi D}\) events.

The misidentification of charged particles is studied by a modification of the selection criteria.
The pion mass is assigned to the track with the charge opposite to \(q_\text{vtx}\) in the kinematic event identification to select \(K^\pm \to e^\pm e^\pm \pi^\mp \gamma\) candidates.
Since this process violates lepton number conservation, all events passing this ``LNV selection'' are considered to be events with misidentified tracks.
A total of 188 events from the full data set pass the LNV selection.
Using the same selection on the MC samples, it is estimated that most of those events are genuine \(K_{2\pi D}\) decays with misidentified \(\pi^\pm\) and \(e^\mp\) tracks, while \(\num{42(18)}\) data events are not accounted for.
The \(x\) distribution of these events is added to the reconstructed \(K_{2\pi D}\) MC one.
The TFF slope shifts by \(\Delta a = \num{0.06e-2}\), which is considered as a systematic uncertainty.

Removing the \(K_{\mu3D}\) and \(K_{e3D}\) MC samples from the fit procedure results in a shift of the slope of \(\Delta a= \num{0.01e-2}\).
%This is considered as a estimate of the systematic uncertainty on the TFF slope due to the neglected \(\pizero\) sources as the other neglected kaon decay modes producing neutral pions have a similar total branching fraction (\SI{3.6}{\percent}).
This is considered as an estimate of the systematic uncertainty on the TFF slope due to the neglected \(\pizero\) sources as the other neglected kaon decay modes producing neutral pions account for less than \SI{4}{\percent} of \(\pi^0\) production.

The acceptance of the \(K_{2\pi D}\) selection for the \(K_{2\pi}\) decay followed by \(\pi^0\to\gamma\gamma\) is estimated with MC simulations to be smaller than \num{e-7}, confirming that the minimal distance requirement between tracks in the first DCH efficiently removes the events with photon conversion.
The reduction of detector acceptance by the Pb bar (\refsec{sample_trigger}) does not lead to any systematic uncertainties since events with a particle within the lead bar acceptance are discarded.

%The expected number of selected events has been estimated usingto be about 160.

%It has been tested in two different ways: randomly
%accept one of the PID hypothesis when both of them are valid, and modify the
% PID to select decays forbidden in the standard model. Since the fraction
%of events rejected with two valid hypothesis is less than 0.1\% and they follow the
%Dalitz \(x\) distribution, the effect is small: \(\Delta a =
%\num{0.01e-2}\). The modified PID assigns a 
%There is no experimental limit on the BR(\(K^\pm \to e^\pm e^\pm
%\pi^\mp \gamma\)), however there is an upper limit on the BR(\(K^\pm \to e^\pm
%e^\pm \pi^\mp \))\( < \num{6.4e-10}\) \cite{PDG_2014}. 
%A total of 188 data
%events and 145.55 MC events are accepted by this selection. Both follow the
%same \(x\) distribution which is not the one of the \(\pizerod\). This leaves
%It is found that 42.55 events are possible background events which have
%been mis-identified. The shift resulting from their addition to the default
% data and MC \(x\) distributions is \(\Delta a = \num{0.06e-2}\).

%Finally, to test the sensitivity of the result to different data taking
%conditions, the data have been divided into groups of sub-samples according to
%the beam configuration, magnet polarity or decay region in the detector. The
%full fitting procedure was run on each of them and the results were compared.
%The TFF slope measured for each sub-sample was compatible within the
%statistical uncertainty with the complementary ones.

%%%%%%%%%%%%%%%%%%%%%%%%%%%%%%%%%%%%%%%%%

\section{Result}
\labelsec{results}
The statistical and systematic uncertainties discussed in the previous sections are summarised in \reftab{error-budget}.
The result of the measurement of the $\pi^{0}$ TFF slope parameter is
\begin{equation*}
    \label{eq:pi0_TFF_final}
    a = \left(3.68 \pm 0.51_{\text{stat}} \pm 0.25_{\text{syst}} \right)
    \times 10^{-2} = \left( 3.68 \pm 0.57 \right) \times 10^{-2},
\end{equation*}
which is in good agreement with the theoretical predictions \cite{Pi0-TFF-Gell-Mann_1961,Pi0-TFF-Lichard_2011,Pi0-radcorr-Kampf_2006, Pi0-TFF-Masjuan_2012,Pi0-TFF-Hoferichter_2014,Pi0-TFF-Husek_2015}.
A comparison with previous \(\pizerod\) measurements is shown in \reffig{world_comparison}.

\begin{table}[H]
    \centering
    \caption{Summary of the uncertainties.}
    \begin{tabular}[H]{|l|c|}
      \hline
      Source & $\Delta a \times10^{2}$ \\
      \hline
      \hline
      Statistical -- data & 0.48\\
      Statistical -- MC   & 0.18\\
      \hline
      \textbf{Total statistical}   & \textbf{0.51}\\
      \hline
      \hline
      Spectrometer momentum scale & 0.16\\
      Spectrometer resolution & 0.05 \\
      LKr calibration & 0.04\\
      Beam momentum spectrum simulation & 0.03\\
      Calorimeter trigger inefficiency & 0.06\\
      Accidental background & 0.15\\
      Particle misidentification & 0.06\\
      Neglected \(\pi^0_D\) sources & 0.01\\
      Higher order radiative contributions &\(<0.01\)\\
      \hline
      \textbf{Total systematic}    & \textbf{0.25}\\
      \hline
    \end{tabular}
    \label{tab:error-budget}
\end{table}

\begin{figure}[ht!]
    \centering
    \includegraphics[width=0.55\columnwidth]{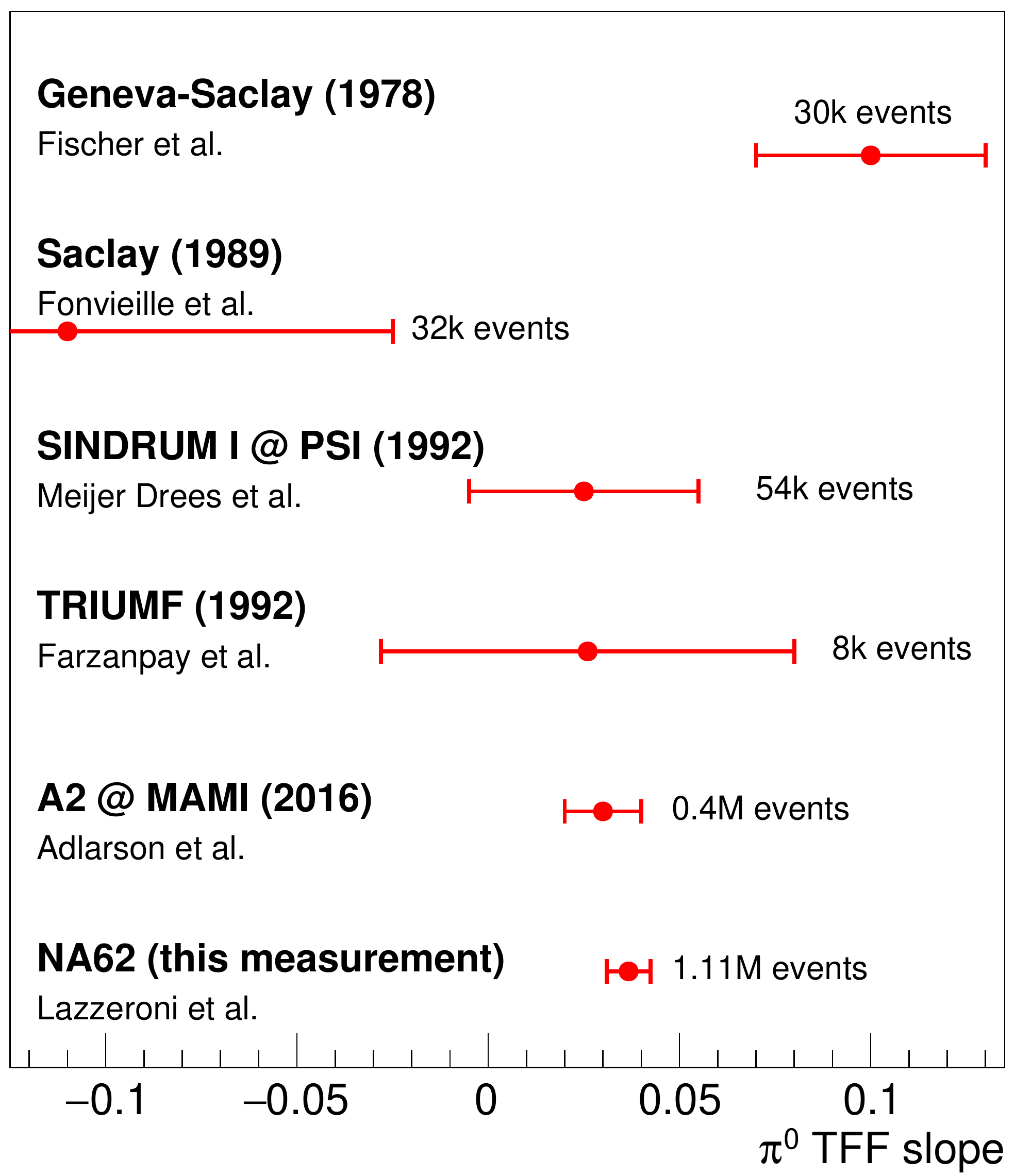}
    \caption{
		Comparison of the \(\pizero\) TFF slope measurements in the time-like momentum transfer region \cite{Pi0-TFF-Fischer_1977, Pi0-TFF-Fonvieille_1989, Pi0-TFF-Farzanpay_1992, Pi0-TFF-MeijerDrees_1992, Pi0-TFF-Adlarson_2016}.
	}
    \labelfig{world_comparison}
\end{figure}

\section*{Conclusions}
\labelsec{world-data}

The slope of the electromagnetic transition form factor of the \(\pizero\) is measured from a sample of \(1.11\times 10^6\) \(\pizero\) Dalitz decays.
The result \(a = \qty(3.68 \pm 0.57)\times 10^{-2}\) represents the most precise measurement of the form factor slope in the time-like momentum region.
The \SI{15}{\percent} relative uncertainty represents an improvement by a factor of 2 with respect to the previous best measurement~\cite{Pi0-TFF-Adlarson_2016}.  

\section*{Acknowledgements}
We express our gratitude to the staff of the CERN laboratory and the technical staff of the participating laboratories and universities for their efforts in the operation of the SPS accelerator, the experiment and data processing.
We thank T. Husek for fruitful discussions and collaboration on the \(\pizerod\) decay generator development.
%\end{linenumbers}

%%%%%%%%%%%%%%%%%%%%%%%%%%%%%%%%%%%%%%%%%%%%

\bibliographystyle{h-physrev}
\bibliography{references}

\end{document}